\documentclass[journal]{IEEEtran}
\usepackage{cite}
\usepackage{bm}
 \usepackage{pgfplots}
\usepackage{multirow}
\usepackage[cmex10]{amsmath}
\usepackage{graphicx}
\usepackage{url}
\usepackage{color}
\usepackage{amssymb}
\usepackage{amsfonts}
\usepackage{amsmath}
\usepackage{extarrows}
\usepackage{graphicx}
\usepackage{amsfonts}
\usepackage{caption}
\usepackage{algorithm}
\usepackage{algpseudocode}
\usepackage{indentfirst}
\usepackage{caption}
\usepackage{esint} 
\usepackage{graphicx}
\usepackage{graphics}
\usepackage{pgfplots}
\usepackage{balance}
\usepackage{tikz}
\usepackage{epsfig}
\usepackage{soul}
\usepackage{caption}
\usepackage{gensymb}
\makeatletter
\def\endthebibliography{%
  \def\@noitemerr{\@latex@warning{Empty `thebibliography' environment}}%
  \endlist
}
\makeatother

\def\bA{{\mathbf{A}}}    
\def\bF{{\mathbf{F}}} \def\bG{{\mathbf{G}}} \def\bH{{\mathbf{H}}}

\def\bU{{\mathbf{U}}}  \def\bW{{\mathbf{W}}}  \def\bY{{\mathbf{Y}}}
\def\bZ{{\mathbf{Z}}}
 \def\bb{{\mathbf{b}}} \def\bc{{\mathbf{c}}}  \def\be{{\mathbf{e}}}
\def\bf{{\mathbf{f}}}    
  \def\bm{{\mathbf{m}}}  
  \def\br{{\mathbf{r}}}  
\def\bu{{\mathbf{u}}}  \def\bw{{\mathbf{w}}}

\def\C{{\mathbb{C}}}

\begin{document}
\title{Leveraging Location Information for RIS-aided mmWave MIMO Communications
\thanks{J. He and M. Juntti are with Centre for Wireless Communications, FI-90014, University of Oulu, Finland (E-mail: jiguang.he@oulu.fi and markku.juntti@oulu.fi).}
\thanks{H. Wymeersch is with Department of Electrical Engineering, Chalmers University of Technology, Gothenburg, Sweden (E-mail: henkw@chalmers.se).}
\thanks{This work is supported by Horizon 2020, European Union's Framework Programme for Research and Innovation, under grant agreement no. 871464 (ARIADNE). This work is also partially supported by the Academy of Finland 6Genesis Flagship (grant 318927) and Swedish Research Council (grant no. 2018-03701).}
}
\author{Jiguang~He,~\IEEEmembership{Member,~IEEE,} Henk~Wymeersch,~\IEEEmembership{Senior Member,~IEEE,} and Markku~Juntti,~\IEEEmembership{Fellow,~IEEE} }
 \maketitle
\begin{abstract}
Location information offered by external positioning systems, e.g., satellite navigation, can be used as prior information in the process of beam alignment and channel parameter estimation for reconfigurable intelligent surface (RIS)-aided millimeter wave (mmWave) multiple-input multiple-output networks. Benefiting from the availability of such prior information, albeit imperfect, the beam alignment and channel parameter estimation processes can be significantly accelerated with less candidate beams explored at all the terminals. We propose a practical channel parameter estimation method via atomic norm minimization, which  outperforms the standard beam alignment in terms of both the mean square error and the effective spectrum efficiency for the same training overhead. 
\end{abstract}

\section{Introduction}
Reconfigurable intelligent surfaces (RISs) are expected to play a pivotal role in the millimeter wave (mmWave) multiple-input multiple-output (MIMO) systems with very-low cost and near-zero power consumption~\cite{Huang2020,di_renzo_smart_2019}. They can be seamlessly incorporated into existing systems and used for maintaining the connectivity when the direct line-of-sight (LoS) path between the base station (BS) and mobile station (MS) encounters blockage, which is commonly seen in mmWave MIMO systems~\cite{Bai2015}. Similar to  point-to-point mmWave MIMO systems, efficient channel state information (CSI) acquisition is challenging~\cite{he2020channel,ardah2020trice,Wang2020}. However, the prior information on the channel parameters, e.g., angles of departure (AoDs) and angles of arrival (AoAs), derived from an out-of-bound location information system (e.g., global positioning system (GPS) for outdoor environment and ultra  wideband  (UWB) for indoor environment) on the MS and environmental objects, can help \cite{alexandropoulos2017position,Garcia2016}. In~\cite{Va2017,Va2019}, the authors considered the beam alignment in mmWave vehicle-to-infrastructure (V2I) communications based on position information. Both supervised offline learning and unsupervised online learning were considered therein. In~\cite{xie2018position}, Xie \textit{et al.} utilized the position information to construct the beams, taking into consideration the position error, and found the best beams with basis pursuit for mmWave MIMO communications. In~\cite{Garcia2019}, Garcia \textit{et al.} studied the synergy between communication and positioning and proposed an in-band position-aided beam alignment protocol, which does not require the involvement of other positioning systems. 

In this letter, we study the effect of prior location information on channel parameter estimation (in particular angular parameters) in RIS-aided mmWave MIMO systems and evaluate the performance in terms of the mean square error (MSE) and the effective spectrum efficiency (SE). Unlike beam alignment approach, the performance will not be limited by the predetermined beam codebooks~\cite{Va2017,Va2019,xie2018position}. Also, unlike randomly designed beams for channel parameter estimation~\cite{wang2021joint}, we harness rough location information for the design of directional training beams, followed by atomic norm minimization (ANM) for channel parameter extraction. Different benchmark schemes, including beam alignment approach, are evaluated to verify the superiority brought by leveraging the prior location information.  

\section{System Model}
\begin{figure}[t]
	\centering
	\includegraphics[width=1\linewidth]{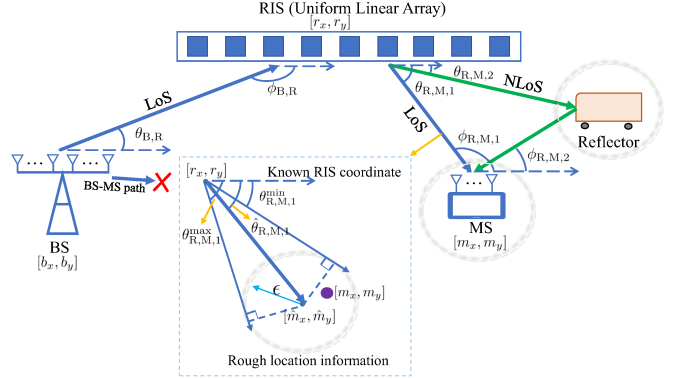}
	\caption{Coarse location information for RIS-aided mmWave MIMO systems with the direct BS-MS path blocked.} 
	\label{System_model}
\end{figure}
The system model is depicted in Fig.~\ref{System_model}, where the direct BS-MS path is blocked and the BS communicates with the MS via the RIS (i.e., BS-RIS-MS path). All the terminals are assumed to equipped with an uniform linear array (ULA), but the extension to uniform planar array (UPA) is feasible. The BS-RIS channel is assumed to have a single LoS path, denoted as $\bH_{\text{B,R}}\in\mathbb{C}^{N_\text{R} \times N_\text{B}}$ with $N_\text{B}$ and $N_\text{R}$ being the number of antennas at the BS and that of the elements at the RIS, respectively. It can be written as
\begin{equation}
\bH_{\text{B,R}} = \sqrt{N_\text{B}N_\text{R}}\rho_{\text{B,R}}  \boldsymbol{ \alpha}_{\text{R}}(\phi_{\text{B,R}} ) \boldsymbol{\alpha}_{\text{B}}^{\mathsf{H}}(\theta_{\text{B,R}} ), 
\end{equation}
where the array response vector $\boldsymbol{ \alpha}_{\text{R}}(\phi_{\text{B,R}} ) \in \C^{N_{\text{R}}\times 1}$ is of the form of
%\HW{The notation is a bit confusing since $\boldsymbol{ \alpha}( \cdot)$ is used to denote different vectors of different length}  
%\begin{align}\label{array_response_vector}
% \boldsymbol{ \alpha}_{\text{R}}(\phi_{\text{B,R}} ) = \frac{1}{\sqrt{N_{\text{R}}}}& [1 \;\;\exp( j \pi \sin(\phi_{\text{B,R}}))  \cdots \nonumber\\ 
%  &\exp( j (N_{\text{R}}-1) \pi \sin(\phi_{\text{B,R}}))]^{\mathsf{T}},
% \end{align}
 \begin{align}%\label{array_response_vector}
 \boldsymbol{ \alpha}_{\text{R}}(\phi_{\text{B,R}} ) = \frac{1}{\sqrt{N_{\text{R}}}}\bigg[1 \;\;e^{ j \pi \sin(\phi_{\text{B,R}})} \cdots
  e^{ j (N_{\text{R}}-1) \pi \sin(\phi_{\text{B,R}})}\bigg]^{\mathsf{T}},\notag
 \end{align}
with half-wavelength inter-element spacing, $\phi_{\text{B,R}}$ and $\theta_{\text{B,R}}$ are the AoA and AoD associated with the BS-RIS channel, $\rho_{\text{B,R}}$ is the propagation path gain, and $j=\sqrt{-1}$ is the imaginary unit. Similar to $ \boldsymbol{ \alpha}_{\text{R}}(\phi_{\text{B,R}} )$, $\boldsymbol{\alpha}_{\text{B}}(\theta_{\text{B,R}} ) \in \C^{N_{\text{B}}\times 1}$ can be formulated in the same manner.

Without loss of generality, we assume that the BS antenna array is parallel to the RIS and no orientation exists at the RIS. Relying on the geometric relationship between the two terminals, we can calculate the angular values, as 
\begin{align}
\theta_\text{B,R} &= \arccos((r_x - b_x)/ \|\br -\bb\|_2),  \label{Geometry_relationship1} \\
\phi_\text{B,R}& = \pi -\theta_\text{B,R} ,\label{Geometry_relationship2}
\end{align}
where $\bb = [b_x ,\; b_y]^{\mathsf{T}}$ and $\br = [r_x ,\; r_y]^{\mathsf{T}}$ are the coordinates of the BS and RIS, respectively.

The RIS-MS channel $\bH_{\text{R,M}}\in\mathbb{C}^{N_\text{M} \times N_\text{R}}$ with $N_\text{M}$ being the number of antennas at MS, is assumed to have multiple resolvable paths in general. Therefore, it is modeled as one LoS path plus a finite number of non-line-of-sight (NLoS) paths, but the NLoS paths are much weaker compared to the LoS path. The RIS-MS channel is in the form of 
\begin{align}
\bH_{\text{R,M}} &=  \sqrt{N_\text{R}N_\text{M}}\bA_{\text{M}}(\boldsymbol{\phi}_\text{R,M}) \mathrm{diag}(\boldsymbol{\rho}_{\text{R,M}}) \bA_{\text{R}}^{\mathsf{H}}(\boldsymbol{\theta}_\text{R,M}),
\end{align}
where $\bA_{\text{M}}(\boldsymbol{\phi}_\text{R,M})  \triangleq  [\boldsymbol{\alpha}_{\text{M}}(\phi_{\text{R,M},1} ), \cdots, \boldsymbol{\alpha}_{\text{M}}(\phi_{\text{R,M}, L_{\text{R,M}}} ) ]$, and $\bA_{\text{R}}(\boldsymbol{\theta}_\text{R,M}) \triangleq  [\boldsymbol{\alpha}_{\text{R}}(\theta_{\text{R,M},1} ), \cdots, \boldsymbol{\alpha}_{\text{R}}(\theta_{\text{R,M}, L_{\text{R,M}}} ) ]$, with $\boldsymbol {\alpha}_{\text{M}}(\cdot) \in \C^{N_{\text{M}}\times 1}$ denoting the array response vector at the MS and $L_{\text{R,M}}$ being the number of paths. The pair of real angles associated with LoS are calculated by following the coordinates of the RIS and MS, similar to~\eqref{Geometry_relationship1}--\eqref{Geometry_relationship2} for the BS-RIS channel. We assume that the orientation is known at the MS and its effect can be easily compensated for with fixed offset angle equal to the known orientation. Other pairs of angles rely on the geometric relationship among the reflecting and scattering points in the environment, the RIS, and the MS.

Finally, we assume to have imperfect \textit{a priori} location information of the MS and environmental objects, e.g., from satellite navigation or indoor positioning technology. We model the prior information on MS location as $\hat{\bm} =[\hat{m}_x,\; \hat{m}_y]^{\mathsf{T}} = \bm + \be$ with $\bm = [m_x ,\; m_y]^{\mathsf{T}}$ being the true coordinate of the MS, where the  estimation error $\be$ is upper bounded as $\|\be\|_2 \leq \epsilon$ (see Fig.~\ref{System_model}). The location accuracy is characterized by $\epsilon\ge 0$. The potential range for the AoD of LoS path in the RIS-MS channel is  
\begin{equation}\label{AoD_range}
\Big[\underbrace{\hat{\theta}_{\text{R,M},1}  - \arcsin(\epsilon/\hat{d}_\text{R,M})}_{\theta_{\text{R,M},1}^{\text{min}}} ; \;\;\; \underbrace{\hat{\theta}_{\text{R,M},1}  + \arcsin(\epsilon/\hat{d}_\text{R,M})}_{\theta_{\text{R,M},1}^{\text{max}}}\Big],
\end{equation}
where $\hat{d}_\text{R,M} = \|\br - \hat{\bm}\|_2$ and $\hat{\theta}_{\text{R,M},1}$ is the estimate of $\theta_{\text{R,M},1}$ based on the prior location information $\hat{\bm}$ and the relationship among $\theta_{\text{R,M},1}$, $\br$, and $\hat{\bm}$, like that in~\eqref{Geometry_relationship1}. Similarly, based on the relationship between $\hat{\theta}_\text{R,M}$ and $\hat{\phi}_\text{R,M}$ in~\eqref{Geometry_relationship2}, the potential range for the AoA of LoS path in the RIS-MS channel is 
\begin{equation}\label{AoA_range}
\Big[ \pi - \hat{\theta}_{\text{R,M},1}  - \arcsin(\epsilon/\hat{d}_\text{R,M}) ; \; \pi-\hat{\theta}_{\text{R,M},1}  + \arcsin(\epsilon/\hat{d}_\text{R,M})\Big].
\end{equation}

The potential ranges for the angular parameters related to the NLoS paths of the RIS-MS channel are calculated in the same manner. These ranges are presumed to be computed at the MS (BS) and broadcast to the RIS and BS (MS).

\section{Location-Based Training Beams}\label{Training_beams}
\begin{figure}[t]
	\centering
	\includegraphics[width=0.8\linewidth]{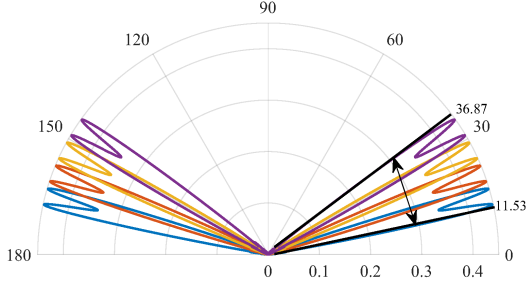}
	\caption{Training beam design based on the prior information on the angular parameters. Each beam is supposed to has uniform response on certain range while null response on the rest. In this example, we use $4$ beams to cover the spatial frequency range $[0.2\; 0.6]$, which is calculated based on the rough location information on the MS or environmental objects.}
	\label{Training_beam_design}
\end{figure}

The training beams used at the RIS and MS are generated based on the prior information on the potential angular ranges \eqref{AoD_range}--\eqref{AoA_range}. Ideally, the group of generated beams used at the RIS should cover the whole potential angular range, where the associated objects (e.g., MS, scatterers, reflectors) may locate. The same principle is applied to the MS. The process is summarized as follows~\cite{Alkhateeb2014}: 
\begin{itemize}
\item Uniformly quantize the spatial frequency $(f = \sin(\theta)$ with $\theta$ being the angular parameter from the individual MIMO channels) range $[-1;\;+1]$ into discrete bins
$\{-1+ \frac{1}{M}, -1+ \frac{3}{M}, \cdots,  1-  \frac{1}{M}\}$, with $M > \{N_\text{R}, N_\text{M}\}$; Construct an over-complete dictionary  $\bA_k = [ \boldsymbol{\alpha}_k(-1+ \frac{1}{M}), \boldsymbol{\alpha}_k(-1+ \frac{3}{M}), \cdots, \boldsymbol{\alpha}_k(1 -  \frac{1}{M})]$ for $k \in \{\text{R},\text{M}\}$. 
\item Determine the spatial frequency range from the available imperfect location information, e.g., $f \in [a; \; b]$, where $-1\leq a \leq b \leq 1$; The prior imperfect location information is first transformed into potential angular range, like these in~\eqref{AoD_range} and~\eqref{AoA_range}, and then transformed into spatial frequency range. 
\item Suppose we want to cover the range $[a; \; b]$ other than the entire spatial frequency range $[-1; \; 1]$ (the case with no location information) with $N$ beams. We solve $\bA_k^\mathsf{H} \bF_k = \bG_k$, for $k \in \{\text{R},\text{M}\}$ for $\bF_k \in \mathbb{C}^{N_k \times N}$, where $\bG_k \in \{ 0,1\}^{M\times N}$ is binary matrix with circular shift property, in order to make the designed beam have unit response in a certain range\footnote{ The number of ``1'' in each column is $\lceil \frac{b-a}{N}/ \frac{2}{M}\rceil = \lceil\frac{(b-a)M}{2N}\rceil$. For instance, for the first column of $\bG_k$, we have $[\bG_k]_{ \lceil(a+1)/\frac{2}{M}\rceil : \lceil(a+1)/\frac{2}{M}\rceil  +\lceil\frac{(b-a)M}{2N}\rceil-1,1} =1$, and the rest of $[\bG_k]_{:,1}$ are 0's.} while null response in the remaining. $\bF_k$ can be obtained by the method of least squares (LS) as $\hat{\bF}_k=[\bA_k]^{\dagger}\bG_k$, with $(\cdot)^{\dagger}$ denoting the matrix pseudo-inverse. The spatial frequency spanned by one beam is roughly $({b-a})/{N}$.
\item Each column of $\hat{\bF}_k$ is regarded as a training beam, used during the channel parameter estimation or beam alignment process. For the RIS, each entry in $\hat{\bF}_\text{R}$ needs to be constant-modulus in order to satisfy the hardware constraint. Thus, we project each generated beam to the constant-modulus vector space. 

\end{itemize}
An example for the designed beams at the MS is depicted in Fig.~\ref{Training_beam_design}. Other possible beam codebook designs~\cite{Wang2020_VT} are left for our future study. 

\section{Channel Parameter Estimation via ANM}

The beamforming gain (or effective SE) will be limited by the resolution of the designed beams (according to Section~\ref{Training_beams}) in the beam alignment process, proven by the results of one benchmark scheme in Section~\ref{Sim_res}. Therefore, we resort to ANM based channel parameter estimation to obtain a refinement (higher-resolution than directly provided by the designed beams) of the channel parameters. This enables a better MS combiner and RIS phase control design compared to the ones chosen from the training beam codebooks. 

\subsection{Observation Model}
During the training process, the beamformer at the BS is designed based on the known coordinates of the BS and RIS, and the BS steers directly towards the RIS.\footnote{For the location and orientation of the RIS, they can be hard-coded to the system upon its deployment and broadcast to the MSs.} That is, the beamformer at the BS is fixed as $\bf =\boldsymbol{\alpha}_{\text{B}}(\theta_{\text{B,R}})$.

Assume $T$ different RIS phase control matrices  $\boldsymbol{\Omega}_t$ (with diagonal elements chosen from the columns of $\hat{\bF}_{\text{R}}$, studied in Section~\ref{Training_beams}) are taken into consideration with $P$ combining beams $\bw_p$ used at the MS (chosen from the columns of $\hat{\bF}_{\text{M}}$, studied in Section~\ref{Training_beams}), the received signal matrix $\bY\in\C^{P \times T}$ during the training phase are summarized as 
\begin{equation}
\bY = \Big[\bW^{\mathsf{H}} \bH_{\text{R,M}} \boldsymbol{\Omega}_1 \bH_{\text{B,R}} \bf s, \cdots, \bW^{\mathsf{H}} \bH_{\text{R,M}} \boldsymbol{\Omega}_T \bH_{\text{B,R}} \bf s\Big] + \bW^{\mathsf{H}}\bZ, 
\end{equation}
where $s$ is the transmitted symbol and set to be $1$,  $\bW = [\bw_1, \cdots, \bw_P]$ contains $P$ beams, exploited at the MS, and $\bZ$ is the additive noise at the MS with each entry following $\mathcal{CN}(0, \sigma^2)$. $\bY$ can be further expressed as 
\begin{align}\label{rec_signal}
\bY &= \sqrt{N_\text{B}N_\text{R}}\rho_{\text{B,R}} \Big[ \bW^{\mathsf{H}} \bH_{\text{R,M}} (\boldsymbol{\omega}_1  \circ \boldsymbol{ \alpha}_{\text{R}}(\phi_{\text{B,R}} )), \cdots,\nonumber\\
&\;\;\;\;\;\;\bW^{\mathsf{H}} \bH_{\text{R,M}} (\boldsymbol{\omega}_T  \circ \boldsymbol{ \alpha}_{\text{R}}(\phi_{\text{B,R}} )) \Big] + \bW^{\mathsf{H}}\bZ,
\end{align}
where $\boldsymbol{\omega}_t=\mathrm{diag}(\boldsymbol{\Omega}_t)$, for $t = 1,\cdots, T$, recall that $\boldsymbol{\omega}_t$ is chosen from the columns of $\hat{\bF}_{\text{R}}$, and $\circ$ denotes the Hadamard product. 

\subsection{Case 1: Only LoS Path from RIS to MS ($L_{\text{R,M}} =1$) }
In this case, the RIS-MS channel has only a LoS path. The received signal associated with one training beam pair (one at the RIS, and one at the MS) can be expressed as 
\begin{align}
[\bY]_{l,k} &= \xi  \bw_l^{\mathsf{H}}\boldsymbol {\alpha}_{\text{M}}(\phi_{\text{R,M}} ) \boldsymbol {\alpha}_{\text{R}}^{\mathsf{H}}(\theta_{\text{R,M}} ) (\boldsymbol{\omega}_k  \circ \boldsymbol{ \alpha}_{\text{R}}(\phi_{\text{B,R}} )) +[\bW^{\mathsf{H}}\bZ]_{l,k}\nonumber\\
& =   \xi   \bw_l^{\mathsf{H}}\boldsymbol {\alpha}_{\text{M}}(\phi_{\text{R,M}} ) \boldsymbol{\omega}_k^{\mathsf{T}}( \boldsymbol {\alpha}_{\text{R}}^*(\theta_{\text{R,M}} )   \circ \boldsymbol{ \alpha}_{\text{R}}(\phi_{\text{B,R}} ))+[\bW^{\mathsf{H}}\bZ]_{l,k} \nonumber\\
& =   \xi   \bw_l^{\mathsf{H}}\boldsymbol {\alpha}_{\text{M}}(\phi_{\text{R,M}} ) \boldsymbol{\omega}_k^{\mathsf{T}}  \boldsymbol {\alpha}_{\text{R}}(\theta_\text{diff}) +[\bW^{\mathsf{H}}\bZ]_{l,k}\nonumber\\
& =   \xi  \bw_l^{\mathsf{H}}\boldsymbol {\alpha}_{\text{M}}(\phi_{\text{R,M}} ) \boldsymbol {\alpha}_{\text{R}}(\theta_\text{diff} )^{\mathsf{T}} \boldsymbol{\omega}_k +[\bW^{\mathsf{H}}\bZ]_{l,k},  \nonumber
\end{align}
where $\xi = N_\text{R} \sqrt{N_\text{B}N_\text{M}}\rho_{\text{B,R}} \rho_{\text{R,M}} $, $\theta_\text{diff} \triangleq \mathrm{asin}( \sin(\phi_{\text{B,R}})-\sin(\theta_{\text{R,M}}) ) $. 
The channel estimation problem is equivalent to LoS channel parameter estimation with prior angular/frequency information. We reformulate the received signal matrix as
\begin{equation}\label{received_signal_matrix}
\bY = \bW^{\mathsf{H}} \tilde{\bH} \bar{\boldsymbol{\Omega}}+\bW^{\mathsf{H}} \bZ, 
\end{equation}
where $\tilde{\bH}$ is a rank-one channel matrix, as
\begin{equation}
\tilde{\bH} = \xi \boldsymbol {\alpha}_{\text{M}}(\phi_{\text{R,M}} ) \boldsymbol {\alpha}_{\text{R}}^{\mathsf{T}}(\theta_\text{diff} ),
\end{equation}
and $\bar{\boldsymbol{\Omega}} = [\boldsymbol{\omega}_1,\cdots,\boldsymbol{\omega}_T ]$. The low-complexity decoupled ANM~\cite{ZHANG201995} is exploited to estimate the AoA of RIS-MS channel and angular difference associated with the RIS. By defining $\bar{\bU} = \tilde{\bH} \bar{\boldsymbol{\Omega}}$ as  $ \bar{\bU}= \boldsymbol {\alpha}_{\text{M}}(\phi_\text{R,M}) \bar{\bc} $ with $\bar{\bc} = \xi \boldsymbol{\alpha}_{\text{R}}^{\mathsf{T}}(\theta_\text{diff}) \bar{\boldsymbol{\Omega}}$, the estimation of $\phi_\text{R,M}$ based on $\bY$ can be formulated as regularized denoising, 
\begin{equation}\label{regularized_denoising}
\min_{\bar{\bU}} \frac{\mu}{2} \|\bar{\bU}\|_{\mathcal{A}_M} + \frac{1}{2}\|\bY-  \bW^{\mathsf{H}}\bar{\bU}\|_{\mathrm{F}}^2,
\end{equation}
where $ \|\bar{\bU}\|_{\mathcal{A}_M}$ is the atomic norm of $\bar{\bU}$, as 
\begin{align}
\|\bar{\bU}\|_{\mathcal{A}_M} & = \mathrm{inf}_{\{\bar{\bu}, \bZ\}}\Big\{\frac{1}{2 T}\mathrm{Tr}(\bZ) + \frac{1}{2 N_{\text{R}}} \mathrm{Tr}(\mathrm{Toep}(\bar{\bu}))\Big\}, \nonumber\\
&\text{s.t.} \;\begin{bmatrix} \mathrm{Toep}(\bar{\bu})  & \bar{\bU}\\
\bar{\bU}^{\mathsf{H}}& \bZ
\end{bmatrix} \succeq \mathbf{0},
\end{align}
with the atomic set as $\mathcal{A}_M = \{ \boldsymbol{ \alpha}(\phi) \bc^{\mathsf{T}} \in \mathbb{C}^{N_\text{R} \times T}: \phi \in [-\pi;\; \pi], \|\bc\| = 1\}$, $\mu$ is regularization parameter, $\mathrm{Tr}(\cdot)$ denotes the trace operation, and $\mathrm{Toep}(\bar{\bu})$ is a Toeplitz matrix with $\bar{\bu}$ being its first row. The optimization problem in~\eqref{regularized_denoising} can be further expressed as 
\begin{align}\label{ANM_Phi}
\{ \hat{\bar{\bu}}, \hat{\bZ}, \hat{\bar{\bU}}\}=&\arg\min_{\bar{\bu},\bZ,\bar{\bU}} \frac{\mu}{2 T}\mathrm{Tr}(\bZ) + \frac{\mu}{2 N_{\text{R}}} \mathrm{Tr}(\mathrm{Toep}(\bar{\bu}))\nonumber\\
& + \frac{1}{2}\|\bY-  \bW^{\mathsf{H}}\bar{\bU}\|_{\mathrm{F}}^2 \nonumber\\
&\text{s.t.} \;\begin{bmatrix} \mathrm{Toep}(\bar{\bu})  & \bar{\bU}\\
\bar{\bU}^{\mathsf{H}}& \bZ
\end{bmatrix} \succeq \mathbf{0}.
\end{align}
The estimation of $\phi_\text{R,M}$ can be obtained by root finding approach based on $\mathrm{Toep}(\hat{\bar{\bu}})$ with $\hat{\bar{\bu}}$ from~\eqref{ANM_Phi}~\cite{he2020channel}. Similarly, we can estimate the angular difference $\theta_\text{diff}$ in the same manner based on $\bY^{\mathsf{T}}$.

\subsection{Case 2: RIS-MS Multi-path Channel ($L_{\text{R,M}} >1$)}
Recall $\bY$ in~\eqref{rec_signal}, the $(l,k)$th entry of $\bY$ is in the form of 
\begin{align}
[\bY]_{l,k}&=\bw_l^{\mathsf{H}}  \sqrt{N_\text{R}N_\text{M}}\bA_{\text{M}}(\boldsymbol{\phi}_\text{R,M}) \mathrm{diag}(\boldsymbol{\rho}_{\text{R,M}}) \bA_{\text{R}}^{\mathsf{H}}(\boldsymbol{\theta}_\text{R,M}) \nonumber\\
&\times \rho_{\text{B,R}} \sqrt{N_\text{B}N_\text{R}} \boldsymbol{\Omega}_k  \boldsymbol{ \alpha}_{\text{R}}(\phi_{\text{B,R}} ) +[\bW^{\mathsf{H}}\bZ]_{l,k} \nonumber\\
&= \xi \bw_l^{\mathsf{H}} \bA_{\text{M}}(\boldsymbol{\phi}_\text{R,M}) \mathrm{diag}(\boldsymbol{\rho}_{\text{R,M}}) \bA_{\text{R}}^{\mathsf{H}}(\boldsymbol{\theta}_\text{R,M})( \boldsymbol{ \alpha}_{\text{R}}(\phi_{\text{B,R}} ) \circ \boldsymbol{\omega}_k)\nonumber\\
&+ [\bW^{\mathsf{H}}\bZ]_{l,k}\nonumber\\
& =  \xi \bw_l^{\mathsf{H}} \bA_{\text{M}}(\boldsymbol{\phi}_\text{R,M}) \mathrm{diag}(\boldsymbol{\rho}_{\text{R,M}}) \bA_{\text{R}}^{\mathsf{H}}(\boldsymbol{\theta}_{\text{diff}} ) \boldsymbol{\omega}_k + [\bW^{\mathsf{H}}\bZ]_{l,k},\nonumber
\end{align}
where $\boldsymbol{\theta}_{\text{diff}}  \triangleq \mathrm{asin} (\sin (\boldsymbol{\theta}_\text{R,M}) - \sin (\phi_{\text{B,R}})\cdot \mathbf{1})$, where $\mathbf{1}$ is an all-one vector.

This is equivalent to rank-$L_\text{R,M}$ (rank-deficient) mmWave MIMO channel estimation, and decoupled ANM can also be applied~\cite{he2020channel}.

\section{Simulation Results}\label{Sim_res}
We evaluate the channel parameter estimation performance and effective SE with prior location information. We set $N_{\text{B}} = N_{\text{M}} = 16$, $N_{\text{R}} = 64$, $\rho_{\text{B,R}} \sim \mathcal{CN}(0,1)$ for all the simulations.\footnote{In practice, an RIS may have hundreds of elements, but during the channel estimation phase, some of them may need to be turned off (i.e., in an absorption mode) in order to achieve an optimal tradeoff between estimation performance and computation complexity.} The signal-to-noise ratio (SNR) is defined as $1/\sigma^2$. For the channel parameter estimation, we introduce the benchmark where $\bW$ and $\{\boldsymbol{\omega}_i\}$ are uniformly designed to cover the entire spatial frequency range $[-1;\; 1]$. For the effective SE evaluation, two different benchmark schemes (beam alignment) are considered: 1) multiple high-resolution beams used at the RIS and MS~\cite{Hur2013}; 2) a single wide beam used at the RIS and MS~\cite{Garcia2016}. For 1), we choose the beam pair associated with the highest received signal power. For 2), a wide beam covering the whole potential spatial frequency range is considered and thus no training overhead is consumed.  The effective SE is expressed as 
\begin{equation}%\label{average_se_bound}
    R \hspace{-0.05cm}=\hspace{-0.05cm}  \mathbb{E} \hspace{-0.05cm}\Bigg[ \frac{T_c-T_t}{T_c}\log_2\hspace{-0.1cm}\bigg(1 +  \frac{|\bw^{\star \mathsf{H}}  \bH_{\text{R,M}} \boldsymbol{\Omega}^\star \bH_{\text{B,R}} \bf |^2}{\sigma^2 }\bigg)\hspace{-0.1cm}\Bigg] \, \text{b/s/Hz}, \notag
\end{equation}
where $T_c$ is the duration of coherence time, $T_t$ is the duration for training, $\bw^{\star}$ and $\boldsymbol{\Omega}^\star$ are the designed MS combiner and RIS phase control matrix based on the parameter estimates~\cite{he2020channel}.

\begin{figure}[t]
	\centering
\definecolor{mycolor1}{rgb}{0.47059,0.67059,0.18824}%
\definecolor{mycolor2}{rgb}{1.00000,0.00000,1.00000}%
\definecolor{mycolor3}{rgb}{0.49020,0.18039,0.56078}%
\definecolor{mycolor4}{rgb}{0.85098,0.32941,0.10196}%
\begin{tikzpicture}[scale=1\columnwidth/10cm]
 
\begin{axis}[%
width=8cm,
height=6cm,
at={(1.15in,0.764in)},
scale only axis,
xmin=-20,
xmax=0,
xlabel style={font=\color{white!15!black}},
xlabel={SNR [dB]},
ymode=log,
ymin=1e-06,
ymax=1e-02,
yminorticks=false,
ylabel style={font=\color{white!15!black}},
ylabel={MSE},
axis background/.style={fill=white},
xmajorgrids,
ymajorgrids,
legend style={at={(0,0)},legend cell align=left, align=left, draw=white!15!black, anchor=south west}
% legend style={legend cell align=left, align=left, draw=white!15!black},
% legend pos=south west
]
\addplot [color=mycolor1, line width=1.5pt, mark size=4.5pt, mark=diamond, mark options={solid, mycolor1}]
  table[row sep=crcr]{%
-20	0.006745824587063\\
-15	0.005225071650572\\
-10	0.003194021456655\\
-5	0.002073650578239\\
0	0.00084479621366\\
};
\addlegendentry{\footnotesize{Angle diff., no loc. info.}}

\addplot [color=mycolor1, line width=1.5pt, mark size=4.5pt, mark=triangle, mark options={solid, rotate=270, mycolor1}]
  table[row sep=crcr]{%
-20	0.005762451068929\\
-15	0.00400484270792\\
-10	0.00253315440563\\
-5	0.001330549606374\\
0	0.000722766331175\\
};
\addlegendentry{\footnotesize{AoA, no loc. info.}}

\addplot [color=mycolor2, line width=1.5pt, mark size=4.5pt, mark=diamond, mark options={solid, mycolor2}]
  table[row sep=crcr]{%
-20	0.00118123319075\\
-15	0.000637735434919\\
-10	0.000345582386015\\
-5	0.0001625026146\\
0	8.6284862641e-05\\
};
\addlegendentry{\footnotesize{Angle diff., with loc. info.}}

\addplot [color=mycolor2, line width=1.5pt, mark size=4.5pt, mark=triangle, mark options={solid, rotate=270, mycolor2}]
  table[row sep=crcr]{%
-20	0.00107975372437\\
-15	0.000535296100871\\
-10	0.000227387374531\\
-5	0.00010936264326\\
0	7.4004586803e-05\\
};
\addlegendentry{\footnotesize{AoA, with loc. info.}}

\addplot [color=mycolor3, dashed, line width=1.5pt, mark size=4.5pt, mark=diamond, mark options={solid, mycolor3}]
  table[row sep=crcr]{%
-20	0.00611581084082888\\
-15	0.00193398919957688\\
-10	0.000611581084082889\\
-5	0.000193398919957688\\
0	6.11581084082889e-05\\
};
\addlegendentry{\footnotesize{CRLB, Angle diff., no loc. info.}}

\addplot [color=mycolor3, dashed, line width=1.5pt, mark size=4.5pt, mark=triangle, mark options={solid, rotate=270, mycolor3}]
  table[row sep=crcr]{%
-20	0.00433276529432931\\
-15	0.00137014068970104\\
-10	0.000433276529432931\\
-5	0.000137014068970105\\
0	4.33276529432931e-05\\
};
\addlegendentry{\footnotesize{CRLB, AoA, no loc. info.}}

\addplot [color=mycolor4, dashed, line width=1.5pt, mark size=4.5pt, mark=diamond, mark options={solid, mycolor4}]
  table[row sep=crcr]{%
-20	0.000663510496629163\\
-15	0.000209820442077763\\
-10	6.63510496629163e-05\\
-5	2.09820442077763e-05\\
0	6.63510496629163e-06\\
};
\addlegendentry{\footnotesize{CRLB, Angle diff., with loc. info.}}

\addplot [color=mycolor4, dashed, line width=1.5pt, mark size=4.5pt, mark=triangle, mark options={solid, rotate=270, mycolor4}]
  table[row sep=crcr]{%
-20	0.000333476693004184\\
-15	0.000105454589647396\\
-10	3.33476693004184e-05\\
-5	1.05454589647396e-05\\
0	3.33476693004184e-06\\
};
\addlegendentry{\footnotesize{CRLB, AoA, with loc. info.}}

\end{axis}

% \begin{axis}[%
% width=5.813in,
% height=4.521in,
% at={(0in,0in)},
% scale only axis,
% xmin=0,
% xmax=1,
% ymin=0,
% ymax=1,
% axis line style={draw=none},
% ticks=none,
% axis x line*=bottom,
% axis y line*=left,
% legend style={legend cell align=left, align=left, draw=white!15!black},
% legend pos=south west
% ]
% \end{axis}
\end{tikzpicture}%
	\caption{Angular difference and AoA of $\bH_{\text{R,M}}$ estimation with $P =4,  T = 16$, $N_{\text{B}}= N_{\text{M}} = 16$, $N_{\text{R}} = 64$, $L_{\text{R,M}} =1$.}
		\label{Parameter_estimation_MSE}
\end{figure}
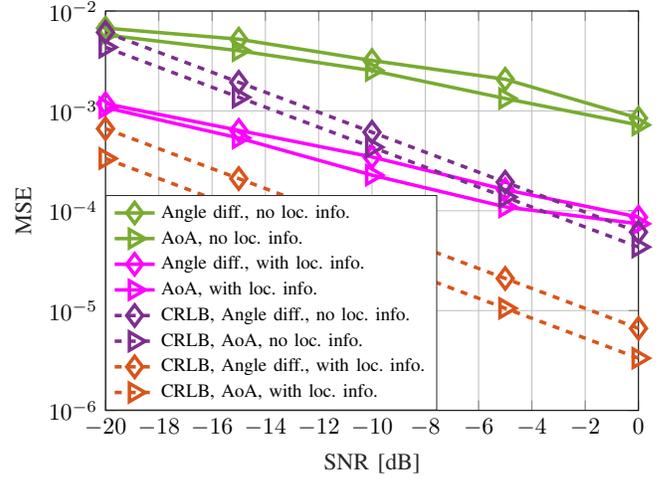

In the first experiment, we compare the channel parameter estimation with prior location information and that without it, where only a LoS path is considered for $\bH_{\text{R,M}}$ and $\rho_{\text{R,M}}\sim \mathcal{CN}(0,1)$. The potential spatial frequency range (one-to-one correspondence with the angular range) determined by the rough location information is $[0.2; \;0.6]$. Without the location information, the training beams are uniformly generated. As shown in Fig.~\ref{Parameter_estimation_MSE}, the performance in terms of MSE of channel parameter estimation (actually in frequency domain other than angular domain) can be improved significantly with the availability of the MS location information, though it is rough. Also, we provide the theoretical Cram\'er–Rao lower bounds (CRLBs) for the angular parameter estimators as a reference for the proposed scheme and the benchmark scheme. As we can seen from the figure, the CRLBs for the proposed scheme are much lower than these for the benchmark scheme thanks to the enhancement on the training beams by the availability of the prior location information.

\begin{figure}[t]
	\centering
\definecolor{mycolor1}{rgb}{0.49020,0.18039,0.56078}%
\definecolor{mycolor2}{rgb}{1.00000,0.00000,1.00000}%
\definecolor{mycolor3}{rgb}{1.00000,0.41176,0.16078}%
\definecolor{mycolor4}{rgb}{0.47059,0.67059,0.18824}%
\begin{tikzpicture}[scale=1\columnwidth/10cm]

\begin{axis}[%
width=8cm,
height=6cm,
at={(0.815in,0.553in)},
scale only axis,
xmin=-20.0460829493088,
xmax=0,
xlabel style={font=\color{white!15!black}},
xlabel={SNR [dB]},
ymin=1.02921413964359,
ymax=11.0292141396436,
ylabel style={font=\color{white!15!black}},
ylabel={Effective SE},
axis background/.style={fill=white},
xmajorgrids,
ymajorgrids,
legend style={at={(0,6cm)},legend cell align=left, align=left, draw=white!15!black, anchor=north west}
%legend style={legend cell align=left, align=left, draw=white!15!black},
%anchor = north west
%legend pos=north west
]
\addplot [color=mycolor1, line width=1.5pt, mark size=4.5pt, mark=diamond, mark options={solid, mycolor1}]
  table[row sep=crcr]{%
-20	4.99059872939272\\
-15	6.30375881596905\\
-10	7.87315080089944\\
-5	9.37469917550684\\
0	10.8765835147018\\
};
\addlegendentry{\footnotesize{Proposed, with loc. info., $T_t = 64$}}

\addplot [color=mycolor2, line width=1.5pt, mark size=4.5pt, mark=triangle, mark options={solid, rotate=270, mycolor2}]
  table[row sep=crcr]{%
-20	4.47053080851802\\
-15	5.70145553000171\\
-10	7.25138901680542\\
-5	8.68418497100018\\
0	10.1735349926187\\
};
\addlegendentry{\footnotesize{Benchmark 1~\cite{Hur2013}, with loc. info., $T_t = 64$}}

\addplot [color=mycolor3, line width=1.5pt, mark size=4.5pt, mark=x, mark options={solid, mycolor3}]
  table[row sep=crcr]{%
-20	1.46291815642138\\
-15	2.3090713298553\\
-10	3.54469917381234\\
-5	4.88787595785456\\
0	6.43786759544451\\
};
\addlegendentry{\footnotesize{Benchmark 2~\cite{Garcia2016}, with loc. info., $T_t = 0$}}

\addplot [color=mycolor4, line width=1.5pt, mark size=3.5pt, mark=o, mark options={solid, mycolor4}]
  table[row sep=crcr]{%
-20	3.14933486259117\\
-15	4.51329326563665\\
-10	6.24500369051865\\
-5	8.1642444840469\\
0	10.1144133035421\\
};
\addlegendentry{\footnotesize{CE via ANM, no loc. info., $T_t = 64$}}

\end{axis}

\end{tikzpicture}%
	\caption{The effective SE with location information for the single-path scenario.}
		\label{effective_SE}
\end{figure}
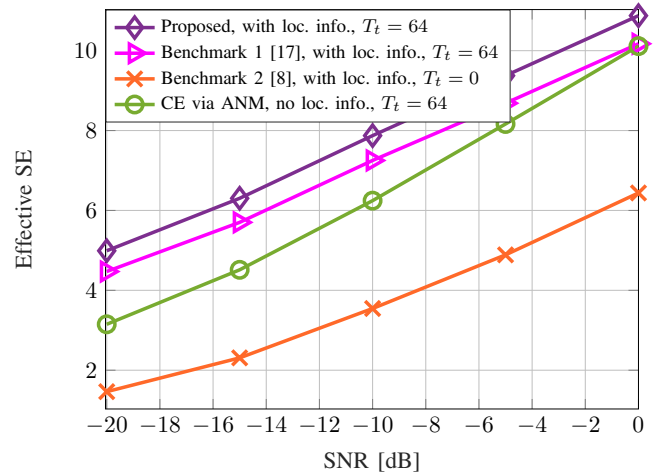

We evaluate the effective SE with only a LoS path for $\bH_{\text{R,M}}$ and $T_c = 500$, taking into consideration of benchmark schemes 1) and 2). As shown in Fig.~\ref{effective_SE}, the performance can be further improved by estimating the channel parameter, based on which a refinement on the MS combiner and RIS control matrix is achieved.\footnote{Throughout the paper, we assume a single RF chain at the MS, so $T_t = PT$. If multiple RF chains are considered at the MS, the duration for training can be further reduced.}

Now, we extend the study to multi-path scenario for $\bH_{\text{R,M}}$ with $\rho_{\text{R,M},1}\sim \mathcal{CN}(0,1)$ and $\rho_{\text{R,M},2}\sim \mathcal{CN}(0,0.1)$, and evaluate the channel parameter estimation and effective SE. In this scenario, the potential spatial frequency ranges associated with the two paths are set as $[0.2; \;0.6]$ and $[0.7; \;0.9]$. From the simulation results in Figs.~\ref{Parameter_estimation_MSE_multi_path} and~\ref{effective_SE_multi_path}, we can observe that the performance can also be improved significantly by taking the rough location information into consideration. 
\begin{figure}[t]
	\centering
\definecolor{mycolor1}{rgb}{0.47059,0.67059,0.18824}%
\definecolor{mycolor2}{rgb}{1.00000,0.00000,1.00000}%
\definecolor{mycolor3}{rgb}{0.49020,0.18039,0.56078}%
\definecolor{mycolor4}{rgb}{0.85098,0.32941,0.10196}%
\begin{tikzpicture}[scale=1\columnwidth/10cm]

\begin{axis}[%
width=8cm,
height=6cm,
at={(0.758in,0.481in)},
scale only axis,
xmin=-20,
xmax=0,
xlabel style={font=\color{white!15!black}},
xlabel={SNR [dB]},
ymode=log,
ymin=1e-05,
ymax=0.0163622189106947,
yminorticks=false,
ylabel style={font=\color{white!15!black}},
ylabel={MSE},
axis background/.style={fill=white},
xmajorgrids,
ymajorgrids,
legend style={at={(0,0)},legend cell align=left, align=left, draw=white!15!black, anchor=south west}
% legend style={legend cell align=left, align=left, draw=white!15!black},
% legend pos=south west
]
\addplot [color=mycolor1, line width=1.5pt, mark size=4.5pt, mark=diamond, mark options={solid, mycolor1}]
  table[row sep=crcr]{%
-20	0.0163622189106947\\
-15	0.0117672266642358\\
-10	0.00719604168601463\\
-5	0.00323539079139365\\
0	0.00170422968817183\\
};
\addlegendentry{\scriptsize{Angle diff., no loc. info.}}

\addplot [color=mycolor1, line width=1.5pt, mark size=4.5pt, mark=triangle, mark options={solid, rotate=270, mycolor1}]
  table[row sep=crcr]{%
-20	0.0128314174723136\\
-15	0.00933373136078902\\
-10	0.0058620504444732\\
-5	0.0028512421975979\\
0	0.00133278843896592\\
};
\addlegendentry{\scriptsize{AoA, no loc. info.}}

\addplot [color=mycolor2, line width=1.5pt, mark size=4.5pt, mark=diamond, mark options={solid, mycolor2}]
  table[row sep=crcr]{%
-20	0.008871807640616\\
-15	0.004579189994482\\
-10	0.002594573903005\\
-5	0.001362514478695\\
0	0.000567538463859\\
};
\addlegendentry{\scriptsize{Angle diff., with loc. info.}}

\addplot [color=mycolor2, line width=1.5pt, mark size=4.5pt, mark=triangle, mark options={solid, rotate=270, mycolor2}]
  table[row sep=crcr]{%
-20	0.00497431355992721\\
-15	0.00281757583072448\\
-10	0.00140457306528446\\
-5	0.000623398444977335\\
0	0.000367967281377882\\
};
\addlegendentry{\scriptsize{AoA, with loc. info.}}

\addplot [color=mycolor3, dashed, line width=1.5pt, mark size=4.5pt, mark=diamond, mark options={solid, mycolor3}]
  table[row sep=crcr]{%
-20	0.00533866130146736\\
-15	0.00168823293688357\\
-10	0.000533866130146737\\
-5	0.000168823293688357\\
0	5.33866130146737e-05\\
};
\addlegendentry{\scriptsize{CRLB, Angle diff., no loc. info.}}

\addplot [color=mycolor3, dashed, line width=1.5pt, mark size=4.5pt, mark=triangle, mark options={solid, rotate=270, mycolor3}]
  table[row sep=crcr]{%
-20	0.00284146109135597\\
-15	0.000898548893143265\\
-10	0.000284146109135597\\
-5	8.98548893143265e-05\\
0	2.84146109135597e-05\\
};
\addlegendentry{\scriptsize{CRLB, AoA, no loc. info.}}

\addplot [color=mycolor4, dashed, line width=1.5pt, mark size=4.5pt, mark=diamond, mark options={solid, mycolor4}]
  table[row sep=crcr]{%
-20	0.0021175023871699\\
-15	0.000669613049430058\\
-10	0.00021175023871699\\
-5	6.69613049430058e-05\\
0	2.1175023871699e-05\\
};
\addlegendentry{\scriptsize{CRLB, Angle diff., with loc. info.}}

\addplot [color=mycolor4, dashed, line width=1.5pt, mark size=4.5pt, mark=triangle, mark options={solid, rotate=270, mycolor4}]
  table[row sep=crcr]{%
-20	0.00122759928706302\\
-15	0.000388200980111803\\
-10	0.000122759928706302\\
-5	3.88200980111803e-05\\
0	1.22759928706302e-05\\
};
\addlegendentry{\scriptsize{CRLB, AoA, with loc. info.}}

\end{axis}
\end{tikzpicture}%
	\caption{Angular difference and AoA of $\bH_{\text{R,M}}$ estimation with $P =6,  T = 24$, $N_{\text{B}}= N_{\text{M}} = 16$, $N_{\text{R}} = 64$, $L_{\text{R,M}} =2$.}
		\label{Parameter_estimation_MSE_multi_path}
\end{figure}
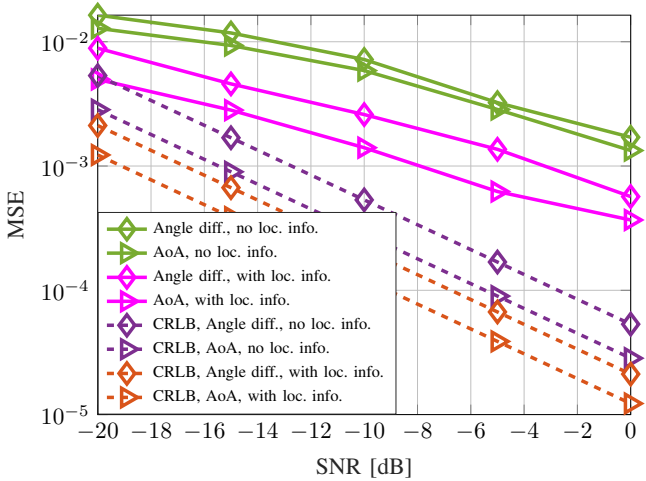
 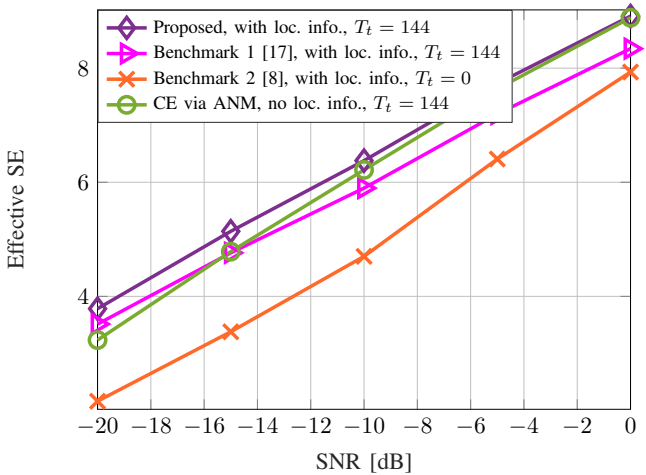
\begin{figure}[t]
	\centering
\definecolor{mycolor1}{rgb}{0.49020,0.18039,0.56078}%
\definecolor{mycolor2}{rgb}{1.00000,0.00000,1.00000}%
\definecolor{mycolor3}{rgb}{1.00000,0.41176,0.16078}%
\definecolor{mycolor4}{rgb}{0.47059,0.67059,0.18824}%
\begin{tikzpicture}[scale=1\columnwidth/10cm]

\begin{axis}[%
width=8cm,
height=6cm,
at={(0.789in,0.504in)},
scale only axis,
xmin=-20,
xmax=0,
xlabel style={font=\color{white!15!black}},
xlabel={SNR [dB]},
ymin=2.02044989775051,
ymax=9.02044989775051,
ylabel style={font=\color{white!15!black}},
ylabel={Effective SE},
axis background/.style={fill=white},
xmajorgrids,
ymajorgrids,
legend style={at={(0,6cm)},legend cell align=left, align=left, draw=white!15!black, anchor=north west}
% legend style={legend cell align=left, align=left, draw=white!15!black},
% legend pos=north west
]
\addplot [color=mycolor1, line width=1.5pt, mark size=4.5pt, mark=diamond, mark options={solid, mycolor1}]
  table[row sep=crcr]{%
-20	3.78377679439809\\
-15	5.13998310480203\\
-10	6.37923573462703\\
-5	7.72968716841029\\
0	8.91143619442856\\
};
\addlegendentry{\footnotesize{Proposed, with loc. info., $T_t = 144$}}

\addplot [color=mycolor2, line width=1.5pt, mark size=4.5pt, mark=triangle, mark options={solid, rotate=270, mycolor2}]
  table[row sep=crcr]{%
-20	3.51331616785851\\
-15	4.76283094809291\\
-10	5.89687187453621\\
-5	7.19572598332968\\
0	8.3412901256275\\
};
\addlegendentry{\footnotesize{Benchmark 1~\cite{Hur2013}, with loc. info., $T_t = 144$}}

\addplot [color=mycolor3, line width=1.5pt, mark size=4.5pt, mark=x, mark options={solid, mycolor3}]
  table[row sep=crcr]{%
-20	2.16228746145832\\
-15	3.37942817398415\\
-10	4.70112281568485\\
-5	6.40692689193377\\
0	7.93156184488\\
};
\addlegendentry{\footnotesize{Benchmark 2~\cite{Garcia2016}, with loc. info., $T_t = 0$}}

\addplot [color=mycolor4, line width=1.5pt, mark size=3.5pt, mark=o, mark options={solid, mycolor4}]
  table[row sep=crcr]{%
-20	3.23398770220169\\
-15	4.78397263153397\\
-10	6.21610241464979\\
-5	7.63510870921673\\
0	8.88299999542059\\
};
\addlegendentry{\footnotesize{CE via ANM, no loc. info., $T_t = 144$}}

\end{axis}
\end{tikzpicture}%
	\caption{The effective SE with location information for the multi-path scenario.}
		\label{effective_SE_multi_path}
\end{figure}

\section{Conclusions}
We have exploited prior information on the location of the MS and environmental objects in RIS-aided mmWave MIMO systems. It has been verified that the introduction of such prior information can further enhance the performance, e.g., MSE, effective SE, compared to the beam alignment approach. 
\balance

\bibliographystyle{IEEEtran}
\bibliography{IEEEabrv,Ref_final}

\end{document}